# An Efficient Framework for Information Security in Cloud Computing Using Auditing Algorithm Shell (AAS)

M. Omer Mushtaq#[1], Furrakh Shahzad#[2], M. Owais Tariq#[3], Mahina Riaz#[4], Bushra Majeed#[5]

[1,2]Department of Computer Science, Pakistan Institute of Engineering and Technology, Multan 60000, Pakistan.
[3,4]Department of Electrical Engineering, Pakistan Institute of Engineering and Technology, Multan 60000, Pakistan.
[5]Department of Information Technology, Bahauddin Zakariya University, Multan 60000, Pakistan.

`[1]omer@piet.edu.pk`, `[2]farrukhshahzad@piet.edu.pk`, `[3]owais@piet.edu.pk`,
`[4]mahinariaz@piet.edu.pk`, `[5]engr.bushra@yahoo.com`

*Abstract*— There is a dynamic escalation and extension in the new infrastructure, educating personnel and licensing new computer programs in the field of IT, due to the emergence of Cloud Computing (CC) paradigm. It has become a quick growing segment of IT business in last couple of years. However, due to the rapid growth of data, people and IT firms, the issue of information security is getting more complex. One of the major concerns of the user is, at what degree the data is safe on Cloud? In spite of all promotional material encompassing the cloud, consortium customers are not willing to shift their business on the cloud. Data security is the major problem which has limited the scope of cloud computing. In new cloud computing infrastructure, the techniques such as the Strong Secure Shell and Encryption are deployed to guarantee the authenticity of the user through 'logs' systems. The vendors utilize these logs to analyze and view their data. Therefore, this implementation is not enough to ensure security, privacy and authoritative use of the data. This paper introduces quad layered framework for data security, data privacy, data breaches and process associated aspects. Using this layered architecture we have preserved the secrecy of confidential information and tried to build the trust of user on cloud computing. This layered framework prevents the confidential information by multiple means i.e. Secure Transmission of Data, Encrypted Data and its Processing, Database Secure Shell and Internal/external log Auditing.

## I. INTRODUCTION

Cloud computing is recognized as one of the huge coming stuffs in information technology (IT) world. Unlike the other old computing systems, cloud computing model offers unlimited arrangement to stock or use the data or programs of client [12]. Cloud computing has made a paradigm shift in the computing industry by delivering computing resources as services over the internet. It has got the name "5th utility" due to its versatile and economic way of making resources available over the internet [1]. Utilities make their resources available to a wider range of customs and charge them only for the usage. Industry analysts and different giant companies such as VMware, Sun, Microsoft, IBM, HP, Google, Dell, Citrix, Amazon and many others have appeared significantly in the favor of cloud computing [16].

Before the beginning of cloud computing in 2000, computing resources such as hardware including processor power, storage and networks bandwidth were purchased and installed in the data centers owned and operated by end users. Cloud computing provides huge economics benefits to its users as accessible resources provided over the internet [2], [18]. Cloud computing is beneficiary for us now a days due to the high bandwidth which was not obtainable in past decades. Although clouds offer more bandwidth and resources for users but still cannot get the attraction of massive users for shifting their work on cloud computing due to security and privacy risk.

Several sensible issues are needed to be resolved to make a cloud-based system more appealing. Cloud computing could be a noisy technology with deep inferences for the internet facilities. Many other problems associated with Service-Level Agreements (SLA), privacy, security and power potency are still needed to be settled. These security issues threaten a lot of potential users, thus keeping them away from leveraging the benefits from this technology, so a correct security model should be designed that assure complete security to the end-user. This kind of models should be able to wipe away all the problems that may occur in a cloud. Every part of the cloud should be analyzed at micro as well as macro level and best solutions should be designed so that multiple users can be engaged and brought on the cloud. Up till that moment, cloud atmosphere is not fully reliable [3]. There are still variety of problems, challenges and implications that are addressed by different researchers, academicians and metal (business intelligence) consultants. Research questions are mentioned below:

How can security and privacy will be achieved in cloud computing? How auditing will be useful in cloud computing for achieving security and privacy? How auditing in cloud computing can enhance the user's trust for shifting their data on cloud? What are improvements in auditing for cloud computing to ensure privacy and security?

The main objective of the research is to develop a cloud computing environment to provide maximum security and privacy to attain more trust of users to shift on cloud. The specific sub objectives are, to develop a secure mechanism for cloud computing which increase the security and privacy for the user in cloud services. A new auditing mechanism in which user can make his own rules for making auditing better. An auditing mechanism in which administrator and user have their own separate auditing protocols and these protocols can be updated by third party. This paper is organized in such manner that Section II describes the effort relating to information security in cloud computing, Section III discusses the proposed methodology of the system, Section IV analyses the experimental results and Section V gives the conclusion.

## II. LITERATURE REVIEW

### A. Cloud Computing Security Challenges

Cloud Computing has been anticipated as the planning of IT industry for the coming generations [22], [23]. It transfers the application programs as well as important databases to the central large data stores [15]. Cloud computing has "special attributes which require risk valuation in the areas like privacy, recovery, data integrity and assessment of the legal concerns in fields such as regulatory compliance, auditing and e-discovery" as Gartner says in [13]. Normally all the security threats related to the cloud computing are magnified by the variety, volume and velocity of the big data [20]. The variations along with cloud based services as well as deployment models lead to the cloud based vulnerabilities and security risks with the basic IT infrastructure [21], [25].

While discussing the security of cloud computing, massive intimidations are raised. Among these threats, researchers mainly focus on the data privacy, integrity and proposed many solutions in order to manage these issues. Priya Metri and Geeta Sarote [2] proposed a model to handle privacy issues in cloud computing. The main issue is the tempering of data, due to which unauthorized modifications can be made in the original data. They proposed multiple solutions to avoid this problem.

Juels et al. [3] introduced PoR (Proof of Retrievability) model in which the retrievability and possession of data files present at the remote file service systems, is ensured by using spot checking and problem correcting codes. Yet, it is not a perfect model as the audit challenges like fixed priority, and public auditability cannot be maintained in the main system. Shacham et al. [4] further made improvements in poor model. They used BLS signatures to build publicly verifiable homomorphic non-linear authenticators. A compact scheme is obtained based on this BLS signatures. But this technique also does not support the auditing, in order to preserve privacy. Sometime it is very difficult for a user of cloud service to know trust level of his/her service provider because of the black box feature of cloud service [17], [19].

Jiang Wang et al. [5] proposed an anonymity based system to deal with confidentiality issues of cloud computing. In this system the data is processed and some information is obscured before it is uploaded on the cloud. When a particular information is desired, the cloud service provider uses this anonymous data and integrates it with the background knowledge that it already has in order to retrieve the desired data. This technique does not require the key management as in the case of traditional cryptography technology. Thus, it is efficient and simple. This approach is not suitable equally for all service providers.

Justus b et al. [6] proposed a privacy stabilizing architecture for database storage. This architecture prevents the outsourced cloud data from both internal and the external attacks. The main elements of architecture are user engine, user interface, rule engine and cloud database. Database is accessed by sending an XML/RPC request through the user interface to user engine, rule engine and finally to cloud database. The privacy is preserved by encrypting the data and conveying the protected identities for the request and its response at all stages. Machine readable access practice rights are also maintained in it. Carrying out the encryption schemes is much easier but there arise some problems in providing the machine readable access rights. If application level safety is the concern of cloud user, then the service provider is basically in charge for physical safety and also for imposing exterior firewall rules. Security for the intermediary layers of application stack is normally shared between the operator and user [14].

While considering the privacy problem of cloud, Miao Zhou et al. [7] proposed another process of access control. Each user's access right is determined by the attributes they are linked with. They presented a two-tier encryption model which consists of two phases (i) base phase and (ii) surface phase. At base phase, the owner of the data encrypts the data to be outsourced. After the initialization is done by the owner of the data, the cloud servers perform the surface phase. In this phase, server re-encryption mechanism is implemented. In this mechanism re-encryption is performed on the encrypted data dynamically, on the request of the owner of the data. Thus, the proposed mechanism preserves the privacy by providing full access control to the data owner while cloud service provider does not have any information about the data.

Rahaman S M. [8] explained PccP (Preserving cloud computing Privacy) model, this is also a privacy preserving model. Base of the model is the consumer layer, in which the request is submitted by the cloud user to have an access to the cloud services. Address Mapping Layer is the second layer in which original IP address related to the access request is modified, thus assuring the privacy of IP address of the data owner. Third and the topmost layer is the privacy preserving layer, in which the privacy check mechanism is implemented for the purpose of preserving the privacy of user's sensitive information. The user can specify the data transparency level and access control in the cloud.

C.Wang et al. [9] proposed another method in which public auditing is carried out on the cloud information. Due to the ubiquitous nature of outsourced data, the traditional cryptographic techniques are not sufficient to achieve full security, so they proposed the concept of third party auditing (TPA). During the process of auditing, TPA cannot gain any knowledge about user's data due to the random masking and Homomorphic authenticators. So, TPA is reliable and capable of acquiring the cloud data storage for the purpose of auditing. Public auditing is performed by two phases and four algorithms. First phase called setup phase consists of the KeyGen and SigGen algorithms. Second phase is the audit phase, which perform the audit process by using GenProof and VerifyProof algorithms. This technique assures the security and privacy preservability of the data in the cloud.

Wang et al.[10] in which the security strength is further improved. They proposed a new protocol for privacy preserving public auditing which achieved zero knowledge leakage also with the improvement in the batch auditing. Proposed design improvement is proved by directing an experiment on the instance of Amazon EC2.

Another technique supported by Boyang Wang et al., this is also a public auditing mechanism known as Oruta, which provides data as well as entity privacy. The approach consists of the three entities: user, TPA and cloud server. Users are organized as original user and group user. The data and its flow is controlled by the original user. Privacy preserving auditing is achieved by Homomorphic Authenticable Ring Structures (HARS) scheme which consists of three algorithms: KeyGen, RingGen and RingVerify [11].

### III.   QUAD SECURITY LAYERS FOR CLOUD TRUSTY ARCHITECTURE

According to the cloud security threats, we have proposed a new concept for data transmission as a secure stream in cloud computing by incorporating quad layers of security in cloud trusty architecture. It is shown in fig 1.

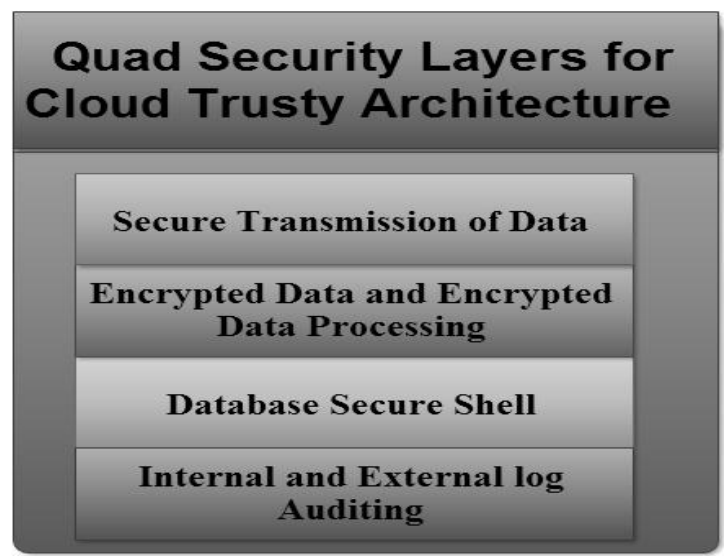

Fig. 1. Cloud Trusty Architecture

The cloud trusty architecture, in our proposed scheme works in the manner shown in fig 2.

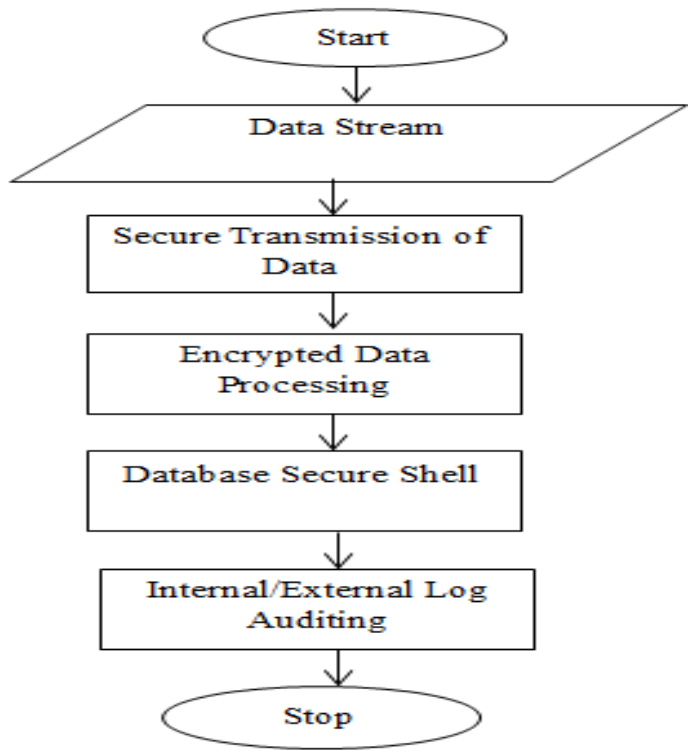

Fig. 2. Flow Diagram of Proposed Technique

### A. Secure Transmission of Data

Initially data is encrypted and then merged with fake data stream to create confusion. It will be merged in different data benches. This generated stream is transmitted to the receiver. On receiver side, first data will be extracted from data benches from specified location and then data stream will be collected. The process at sender and receiver side is shown in fig 3 and fig 4.

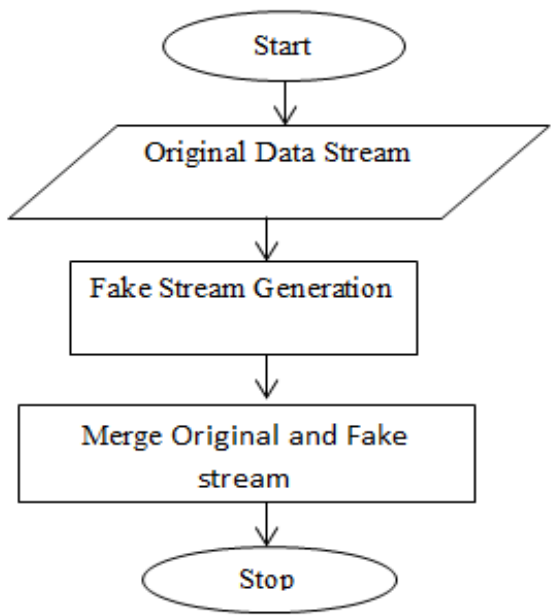

Fig. 3. Transmission Process from Sender Side

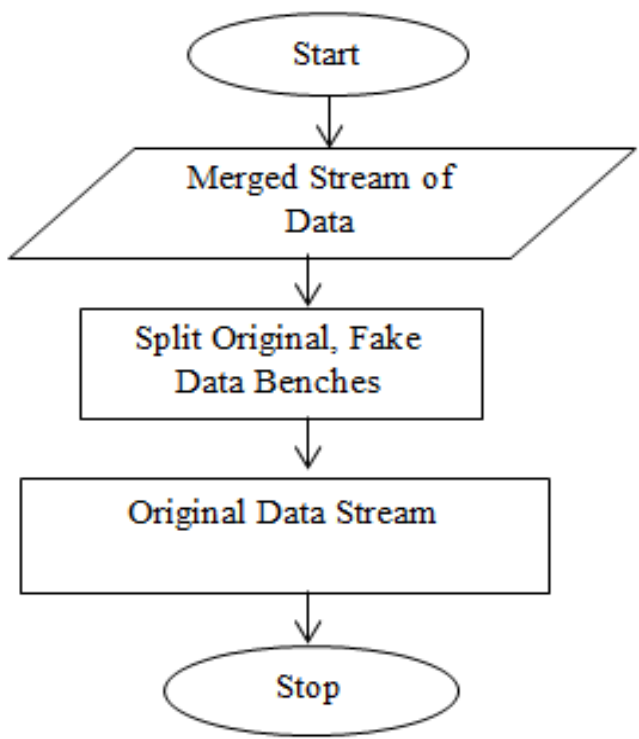

Fig. 4. Data Extraction Process at Receiver Side

The fake data benches and original data benches merging ratio will be random. This depends on the fake data merging ratio. The fake data merging can be set in different ranges. This algorithm will create fake data benches according to given ratio. This ratio can be set priority wise. This priority is based upon the transmission time and security level. If transmission time is preferred then less fake data merging ratio will be added. This will increase the transmission efficiency but data security will be compromised. If security level is preferred then fake data benches merging ratio will be increased but it will add some delays in transmission due to increase of stream size, this will add confusion for hackers over internet during transmission. The major concept of this security layer is to increase the confusion through fake data streaming. This fake data streaming adds confusion for hacker at transmission level. Because many times, users do not know about data hacking due to passive attack, where attacker only observe data during transmission. As data is not changed in passive attack so user doesn't know about data breaches. We merged original data with fake stream during transmission of data so hacker cannot get original data due to huge confusion in it.

By using this technique we can secure our data in transmission state over Internet in cloud computing. Even if the data will be hacked, the hacker doesn't know about accurate data because of different data bench location mixed with false data. Secondly, he doesn't know about the encryption technique used in encryption process. This combination secures the stream while its transmission over the cloud.

### B. *Encrypted Data and Encrypted Data Processing*

Second threat is data breach that meets at cloud side, this data breach occurs during processing and from storage area by hacking. In this technique all the data on server side will remain in encrypted form. All the processing will also be done in encrypted form. The encrypted information after processing will be sent to client. During transmission to the client, the encrypted data again merge with false data and this stream will be sent to the client. This received data will be extracted and real data benches will be collected to make new stream. After making new stream, this will be decrypted and real data will be presented to the user. Decryption will be done only at the client side. The data will remain in encrypted form on cloud side during all over the time. This is explained in fig 5.

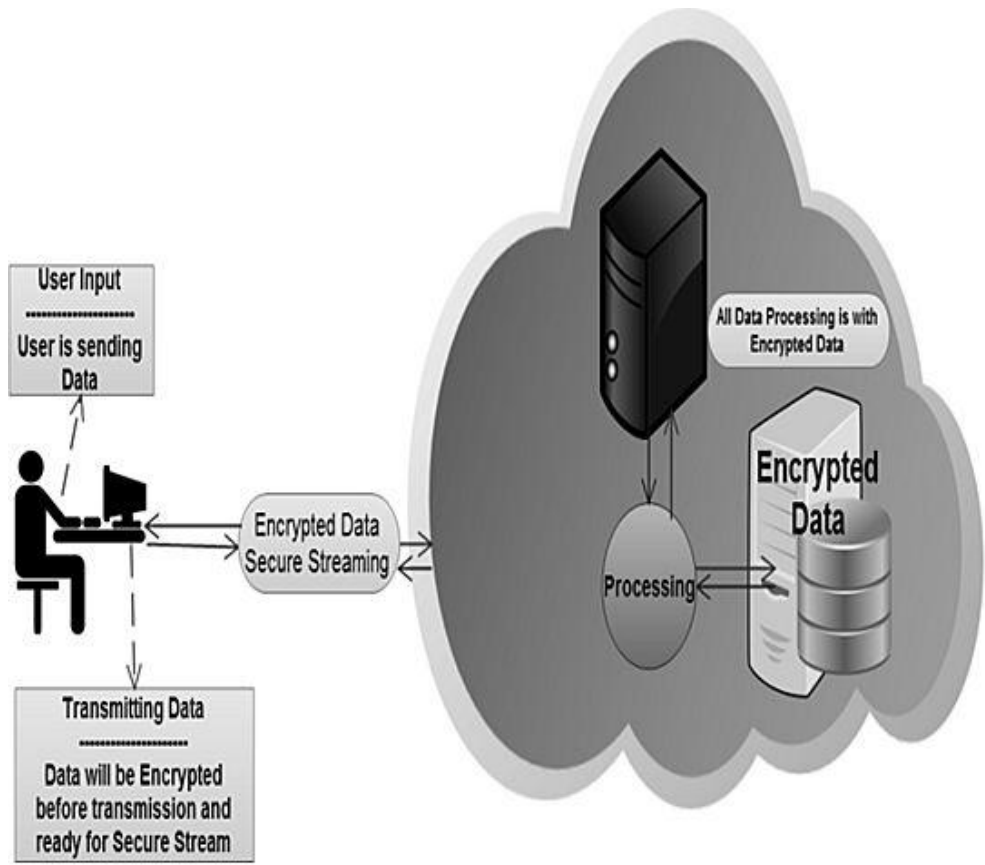

Fig. 5. Encrypted Data Processing

In this manner the user is sending data at client side, the data will be encrypted and merged with fake data to make secure stream for transmission. This secure stream is received at cloud side. After receiving this stream, the original encrypted data will be extracted. This encrypted data will be processed with encrypted data stored in database. After processing, this encrypted data will again merged with fake data and this secure stream will be sent to the client. At client side the secure stream will be received and original data will be extracted. The extracted data will be finally decrypted and presented to the user.

*C. Database Secure Shell*

Third threat for data breaches occur at storage level due to multiple reasons. Like, Identity Theft, Hackers Attempt, Admin or Cloud Provider access their own client's data etc. In this technique we are presenting a security layer named as Secure Shell. All the communication with database will be done with this secure shell. All the data servers will communicate with database by this security layer (Secure Shell). This secure shell will generate logs with each attempt to access data by any data server. It will generate logs by collecting several information stamps of requester. Like, Time Stamp, Identity Stamp, Network Address Stamp, Location Stamp etc. These logs will save in log database for auditing. Data base secure shell is presented in fig 6.

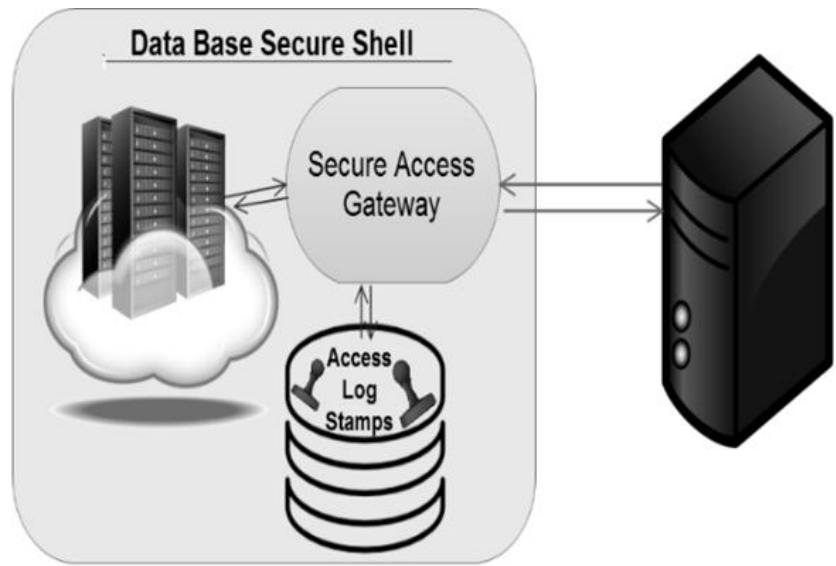

Fig. 6. Database Secure Shell

In this manner the Server will communicate for data access through database with the help of Secure Access Gateway. The Secure Access gateway will keep record of each query for data by using different Log Stamps like Time Stamp, Identity Stamp, Network Address Stamp, Location Stamp, Application Stamp etc. The Secure Access Gateway gets the request from Server and according to this request, the data will be provided to the server. Each time the server requests for any query from database, the Access Log Stamps will be maintained. This Access Log Stamps will be used for auditing in next layer. This will increase the data security, because only Access Logs Stamps are presented for Auditing.

*D. Internal and External Log Auditing*

Forth Layer is built for gaining the user trust, because auditing is used for analysis and finding the irregularity in data access all over the world in all over the fields. For example many companies conduct audit for satisfying their shareholders. Some of researchers have presented auditing in cloud computing for gaining trust of user as well as to rectify their system. They suggested the Third party auditors for auditing. A big threat for data breaches arises here, because all of them suggesting the real data for auditing. Internal and external auditing of our proposed scheme is described in fig 7.

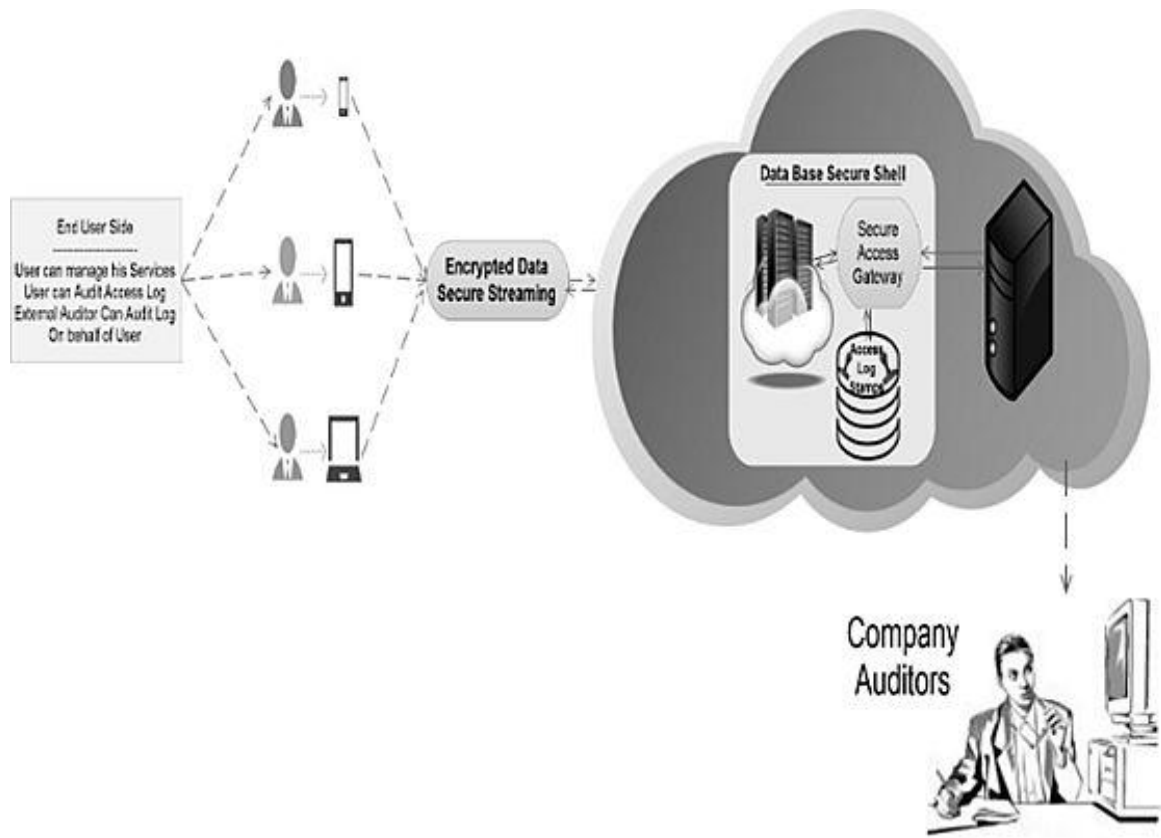

Fig. 7. Internal/ External Log Auditing

- **Internal Auditing**

In Internal Auditing the organization will perform auditing by their appointed or hired auditors. There are two types of such auditing. One is Interim audit and second is Final Audit. The Interim audit will be conducted regularly and after a specific time span. Like after every Quarter (3 Months). The final audit will be conducted annually.

- **External Auditing**

In external audit the user will audit his logs by himself or any hired auditors. User can inspect activities on the basis of logs that either the data is accessed by authorized person. The user can realize the data breaches on the basis of access time, access location and access network. This will make cloud computing more reliable and trustworthy for any user. As the cloud is much facilitated service but user doesn't rely on cloud computing due to data breaches. This fact is affecting on growth of cloud computing.

## IV. DISCUSSION AND ANALYSIS

This portion provides the comparison between standing systems and proposed solution. This portion further describes proposed solution to show how it helps to overcome the limitations and problems of the existing systems.

In this paper, we discussed the security matters and challenges at numerous levels in the cloud computing. Security challenges in cloud computing have high impact to limit the scope of this domain. Due to privacy, integrity and confidentiality concern of data, there is a need of security architecture that should overcome the security risks in cloud computing and reduce the fear of enterprise customer to adopt cloud [24]. Thus, we developed an architecture known as cloud trusty architecture. It provides four defensive layers that ensure security of the outsourced data in the untrusted cloud data center.

The process of implementing or updating the security architecture of Cloud Computing is difficult or even impossible to navigate because of the rapid change and agile nature of information technology. System networks are now evolving very fast with the revolution in the cloud computing. Furthermore, old models for the network security have been proved useless against the new coming malwares and advanced threats. Target data breach is an example, which resulted the loss of more than 100 million user archives. Big multinational companies like Adobe and eBay emphasize on the serious risks associated with the insufficient cloud security.

In addition, a set of roughly integrated security applications and devices make it problematic for the IT groups to identify the threats, as they are incapable to correlate the different security reports, logs, alerts and events. The analysis of our work ensures maximum security for data breaches. It is an integrated framework to provide security for IT organizations. The practical scenario of our scheme is shown fig 8 & fig 9. It discusses the case for 1 byte data bench.

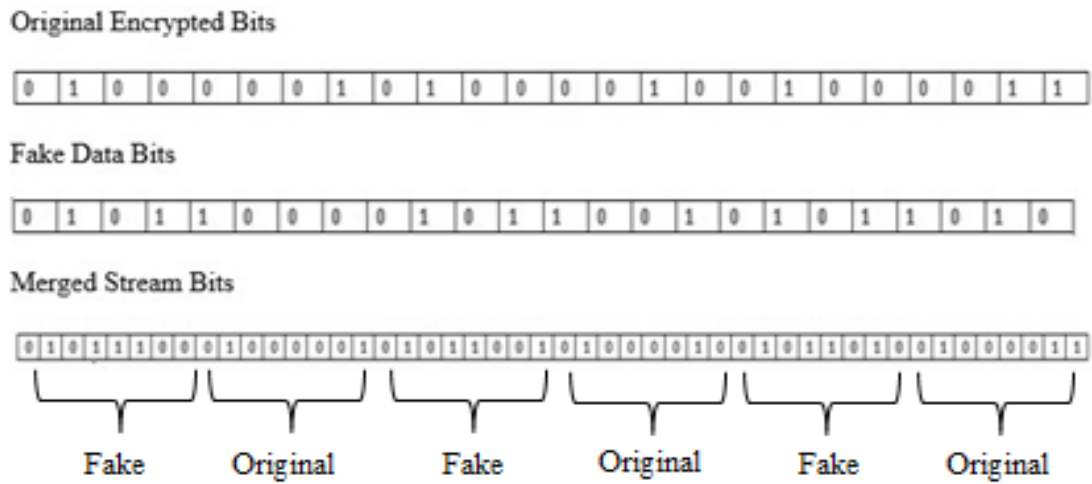

Fig. 8. Noise addition in Data

Noise removal and extraction of original data, in our proposed scheme is as explained in fig 9.

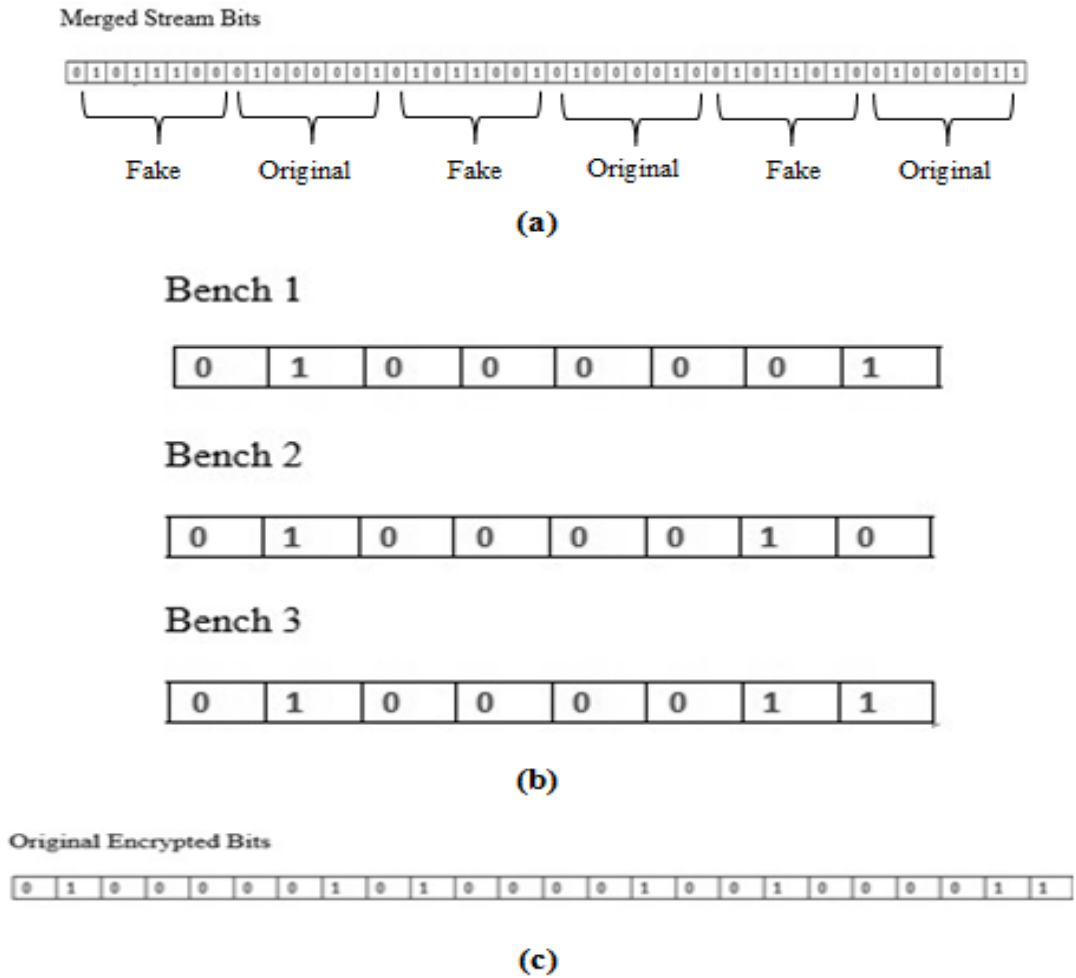

Fig. 9. Original Data Extraction (a) Merged Stream Bits (b) Bench 1, Bench 2 and Bench 3 (c) Original Encrypted Bits

The merged stream of data is encrypted and third defensive layer maintains access logs in the following manner described in fig 10.

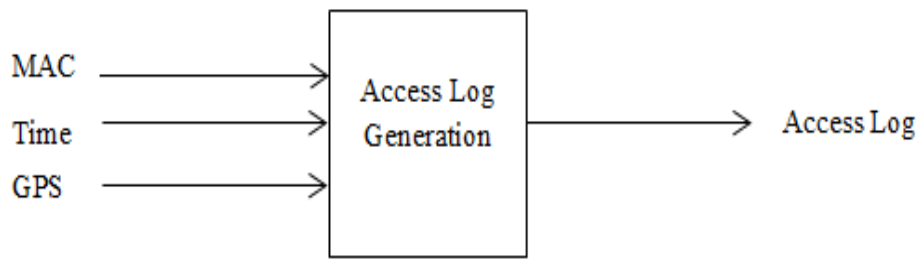

Fig. 10. Access log Generation

Database maintains these access logs for further references. It allow user to view how many times his data is accessed and from where. It prevents passive attacks. Its comparison with some of the famous cloud service provider is discussed in table 1.

Table 1

COMPARISON OF QUAD CLOUD TRUSTY ARCHITECTURE VS GDRIVE AND DROPBOX

| Features | Quad Cloud Trusty Architecture | gDrive | Dropbox |
|---|---|---|---|
| Passive Attack Detection | Yes | No | No |
| Access Log Generation | Yes | No | No |
| Trusted Third Party access | Encrypted data | Original Data | Original Data |
| Internal Auditing | Yes | No | No |
| External Auditing | Yes | Yes | Yes |
| Data Over Cloud | Encrypted Form of Data | Original Data | Original Data |

The framework of our scheme ensures that, data to be transmitted over the cloud is secure enough, as first defensive layer of this architecture stores the data over cloud by merging it with some noise bits. The data over the cloud is noisy for an intruder. Meaningless data stream to confuse an intruder is basically acts as a layer of secrecy in our system. Layer two of this architecture encrypts the data with some encryption algorithms. This creates confusion and diffusion in the data over cloud.

Database secure shell records the access of data by different stamps like time stamp, network address stamp, location address stamp, MAC stamp, etc. Server can request data through secure access gateway, and log stamp is maintained for it. For auditing purposes, these access log stamps are given, not the original data. It ensures that third party can never get access to user's original data. Final layer of this architecture allows the user to audit his data, where he can identify masquerade by viewing his access log details. The user can realize the data breaches on the basis of access time, access location, access network, etc. This will make cloud computing more reliable and trustworthy for any user. There is some computational overhead at server side as the four defensive layers are to be applied on data to shield it. The confidentiality, passive attack detection and integrity of data in this scheme are more esteemed than this overhead.

V. CONCLUSION

The previous work on auditing systems only audit data by the third party auditor. This method usually increase the security risk, as they provide the original data for auditing to the third party. If the third party auditor make illegal use of this data that was presented for auditing, it caused high security risk. In our proposed auditing system, the whole data will be secured in the security shell, where only data logs will be provided to the third party auditor or internal auditor. The original data will not be provided for auditing. This increases data integrity and confidentiality. The data over cloud is also in encrypted form. Moreover, this scheme ensures user that every attempt to access the data can be viewed by access logs, which provides information about location of access, MAC and other log stamps. This architecture provides complete protection on data security, data privacy and data breaches on cloud computing. So it is an effective architecture for gaining user trust over cloud.